\documentclass[preprint2]{aastex63}

\begin{document}

\title{A Pan-STARRS1 Search for Planet Nine}

\correspondingauthor{Michael Brown}
\email{mbrown@caltech.edu}

\author[0000-0002-8255-0545]{Michael E. Brown}
\affiliation{Division of Geological and Planetary Sciences\\
California Institute of Technology\\
Pasadena, CA 9125, USA}

\author[0000-0002-1139-4880]{Matthew J. Holman}
\affiliation{Center for Astrophysics \\
Harvard \& Smithsonian \\
60 Garden Street, Cambridge, MA 02138, USA}

\author[0000-0002-7094-7908]{Konstantin Batygin}
\affiliation{Division of Geological and Planetary Sciences\\
California Institute of Technology\\
Pasadena, CA 91125, USA}

\begin{abstract}
We present a search for Planet Nine using the second data release of the Pan-STARRS1 
survey. We rule out the existence of a Planet Nine with the characteristics
of that predicted in Brown \& Batygin (2021) to a 50\% completion depth of $V=21.5$.   This survey,
along with previous analyses of the Zwicky Transient Facility (ZTF) and Dark Energy
Survey (DES) data, rules out 78\% of the Brown \& Batygin parameter space. Much 
of the remaining parameter space is at $V>21$ in regions near and in the area where the northern galactic plane crosses the ecliptic. 
\end{abstract}

\section{Introduction}
Speculation about the existence of planets beyond the orbit of Neptune began almost as
soon as the announcement of the discovery of Neptune itself \citep{Babinet1848}
and has continued to the present day. While \citet{1993AJ....105.2000S} demonstrated
that no evidence exists in the planetary ephemerides for any significant
perturber, the concurrent discovery of the large population of small bodies in the
Kuiper belt beyond Neptune led to renewed scrutiny of dynamical 
signatures of perturbation in this population. The first concrete hint
of the need for an external perturber -- at least at some point in solar system
history -- came from the discovery of Sedna, with a perihelion at 76 AU, well
beyond where it could have been perturbed by the known planets \citep{2004ApJ...617..645B}. As more objects with the extreme semimajor
axis of Sedna were discovered, they were suggested to be anomalously
clustered around an argument of perihelion of zero, though the physical mechanism
for the apparent clustering was unclear \citep{2014Natur.507..471T}.
Subsequent analysis showed that the apparent clustering in argument
of perihelion is actually a consequence of simultaneous
clustering  in longitude of
perihelion and in pole position, a phenomenon that can be
naturally explained by the presence of a massive planet on 
a distant, eccentric, and inclined orbit \citep{2016AJ....151...22B}. The 
prediction of the existence of this planet has been called 
the Planet Nine hypothesis. Planet Nine is now seen to be capable of accounting for
a range of additional otherwise unexplained phenomena in the solar system, 
including the existence of 
highly-inclined Kuiper belt objects and the existence of retrograde Centaurs \citep{2016ApJ...833L...3B}.

Since the prediction of the existence of this planet, discussions
of alternative explanations have included suggestions 
that it is instead a primordial black hole \citep{2020PhRvL.125e1103S}, 
that the observed effects are caused by the presence of a distant
unseen ring of material \citep{2016MNRAS.457L..89M,2019AJ....157...59S}, or that the planet is instead
a collection of condensed dark matter \citep{2016Ap&SS.361..230S},
though a planet remains a far simpler explanation than these exotic possibilities.
Suggestions
that observational bias might be responsible for the 
observed clustering effects have been made from analysis of limited data sets \citep{2017AJ....154...50S, 2021arXiv210205601N}, 
but analysis of the largest data sets has repeatedly found
the probability of such bias small \citep{2019AJ....157...62B, 2021AJ....162..219B}. While the existence
of Planet Nine remains the
most satisfactory explanation for a range of phenomena, true 
detection of the planet will be required to firmly discount 
these or other alternatives.

The Planet Nine hypothesis makes distinct predictions about the properties of
the planet and its orbit, based on the currently observed distant eccentric Kuiper belt population and on an assumed source population for these bodies.
Under the assumption that the currently observed distant eccentric 
population is sourced from an initial extended disk (rather than, i.e. from 
the inner Oort cloud, \citet{2021ApJ...910L..20B,2023Icar..40615738N}),
\citet[hereafter BB21]{2021AJ....162..219B} use a suite of numerical models corrected 
for observational bias to construct statistical distributions of the mass and
orbital elements of the hypothetical Planet Nine consistent
with the observed clustering. 
All elements are well constrained except
for the mean anomaly; the current observations show only the orbit of Planet Nine,
not the position within the orbit. To aid in the search for Planet Nine and to better understand search limits for different surveys, \citet[hereafter BB22]{2022AJ....163..102B}
construct a synthetic population by sampling from the posterior of the BB21
mass and orbital element distributions. This synthetic population
can be injected into any data set to give a statistical 
representation of Planet Nine parameter space and determine
which parts of parameter space a survey can rule out.

The first wide-field search for Planet Nine examined three years of
the Zwicky Transient Facilities (ZTF) archives (BB22). This search covered most
of the predicted path of Planet Nine (with the exception of regions below
-25 in declination and in the densest regions near the galactic center) and
concluded that a Planet Nine candidate with the predicted parameters
was not detectable in the archive. Though typical ZTF 
images reach a depth of only $r\sim20.5$, injection of the synthetic population into the ZTF
catalog showed that ZTF was sensitive to 56\%  of the predicted parameter space
of Planet Nine.  

The Dark Energy Survey (DES) covered only a modest
fraction of the predicted orbital path of Planet Nine,
but the entire survey region was surveyed for moving objects
with distances from 30 - 2000 AU \citep{2022ApJS..258...41B}. No Planet Nine
or other distant object candidates were found. \citet{2022AJ....163..216B}
used the previously developed 
synthetic population to show that the DES ruled out 10.2\% of the 
predicted phase
space of Planet Nine, of which 5.0\% had not previously been
ruled out by ZTF. Between the ZTF and DES survey 61.2\% of the
predicted Planet Nine phase space has been ruled out.

Here, we extend the search for Planet Nine to the Pan-STARRS1 DR2 (PS1) data base. PS1 observed a similar amount of sky as ZTF, but
with deeper -- though less frequent -- coverage. 
For much of the sky, PS1 should extend the magnitude limit
of the search for Planet Nine by approximately a magnitude.

\section{The Pan-STARRS1 data}
The Pan-STARRS1 survey covered the approximately 3$\pi$ steradians of the sky 
north of a declination of -30$^\circ$. Each area in the sky was covered
approximately 12 times from 2009 to 2015 in each of 5 broadband filters ({\it grizy}
reaching a single epoch depth of approximately 22.0, 21.8, 21.5, 20.9, 19.7,
respectively). 
If Planet Nine was detected by PS1, it would appear as a single night transient 
in each detection. To search for Planet Nine, we
will search for collections of single night transients
which appear at locations consistent with a Keplerian
motion moving  on an orbit within the range of
parameters predicted by \citet{2021AJ....162..219B}.

Directly querying the PS1 catalog for single night transients
is not possible, so  
to construct our list of single night transients,
we begin by
using the PS1 CasJobs SQL server\footnote{http://mastweb.stsci.edu/ps1casjobs/} to download
every detection of every cataloged object for which there
were fewer than 12 detections, with at least one of those
detections in the $g$, $r$, or $i$ bands which would be most
sensitive to a solar-colored Planet Nine. Downloading these
data in small batches required several months of continuous
querying of the server. 
Many of these collections of up to 12 detections
will be real stationary astrophysical
sources which appear at the same location on multiple 
nights. Thus, after downloading,
we discarded any object for which the detections of a 
cataloged object occurred 
on more than a single date. Over the Planet Nine search region --
defined as the declination range over which 99\% of the Planet
Nine synthetic population occurs -- we find 1.26 billion single night
objects, most of which consist of a
single
detection in a night, but a small
number consisting of between
2 and 11 detections at a single location
in one night.

The vast majority of single night
transients in PS1 at the faint end are not real astrophysical 
sources, but rather arise from systematic noise \citep{2016arXiv161205560C}. Our goal is to
find any set of transients that appear to follow
a Keplerian orbit consistent with that
predicted for Planet Nine
over the five year period of the data. 
Finding such a set in the background of a billion
bad detections
is formidable problem.
We thus explore ways to remove at least
some of these bad detections. Any method which 
removes bad detections has the possibility of 
removing real detections also, thus we develop
a method of calibration to take this possibility 
into account in the next Section.

To better understand the characteristics of real
single night transients, we extract any detections
within the data set 
of the first 501090 numbered asteroids 
by selecting every
detection within 2 arcseconds and 2 magnitudes 
of an asteroid's predicted position and brightness 
at each moment of observation. We have high confidence 
that those detections are nearly all real, as chance
alignments on the scale of 2 arcsecs on 
individual PS1 exposures are rare, despite 
the abundance of spurious detections. This sample
probes the full magnitude range of potential PS1
detections including the faintest end where
the asteroids detected are those which were
initially discovered at closer distances 
and were thus brighter but are now more distant and
can have brightness at and well beyond the PS1
magnitude limit.

We examine these 6.21 million asteroid detections
detections and find that we can use two parameters
extracted from the PS1 database, PSFCHI2 and PSFQF {(measures of the
fit of the detection to the PSF and of the total
coverage of the detection to the PSF)}, to help eliminate 
some of the most likely false positives. We make a simple cut
and include only detections with PSFQF greater than 0.99
and, for objects with a magnitude less than 16, PSFCHI2 between
0 and 8, and for fainter objects, PSFCHI2 between 0 and 1.6.
We also reject all objects with a reported magnitude fainter
than 22.5 as this magnitude is both beyond the 
stated depth of PS1 and our asteroid detections 
quickly drop before this magnitude. These simple cuts reduce the number of objects 
detected as single night transients to 772 million.

Significant spatial structure still exists in the locations
of the remaining objects. A plot of the position of each
object on the imaging array shows that each of the 60 
individual detectors has distinct regions with greatly 
increased numbers of detected objects. We create individual
masks for each of the 60 detectors by hand, and discard all
objects within these masked regions. This masking
reduces the number of objects under consideration to 428 million.

Another source for large number of clustered objects is
scattered light from bright stars. We use the ATLAS REFCAT2 star 
catalog
\citep{2018ApJ...867..105T} to examine the positions of all bright stars in the 
data. Stars fainter than about $r\sim 13$ 
have limited issues with scattered light, but brighter
stars are surrounded by clusters of objects. The radius
of these clusters increases with the brightness of the star.
We empirically define a bright star exclusion radius, $r_e$,
based on $m_r$, the $r$ magnitude in REFCAT2, as
$$
r_e=50+0.08(13-m_r)^{4} 
$$
where $r_e$ is in arcseconds. 
After excluding these objects, 314 million objects remain.

Even with detector masking and regions around bright stars
excluded, clusters of objects can sometimes be found
associated with individual image frames. We find that these
can be effectively identified by using a clustering algorithm.
Specifically, we remove all collections of 5 or more 
objects occurring within 40 arcseconds of each other. 
This final cut reduces our data set to 244 million objects.

We have reduced our original data set from 1.2 billion
to 244 million objects, 
a decrease of 80\%. While searching for Planet Nine through 
this large data set remains a formidable task, we could find
no additional filters that appeared to safely further reduce
the data set.
Any one of these cuts in the data has the possibility of 
removing real detections of Planet Nine. Our calibration 
method must by necessity take this possibility into account.

\section{Calibration}
The footprints and depths of individual PS1 pointings are not easily
available, thus we use the method developed in BB22 to self-calibrate the data
set. In short, we use the asteroids identified above as probes of both
the coverage and depth of PS1 on individual nights.

We use the JPL Horizons system\footnote{ssd.jpl.nasa.gov} to calculate the
positions and magnitudes of the first
501090 numbered asteroids
for each night of the 5 year  PS1 survey. We interpolate the positions
to the time of each image taken during a night and keep a log of which
asteroids are and which are not detected each night and of what
their predicted $V$ magnitude for that night was. 
We record this information in $\sim$1.8 square degrees patches of the
sky using an NSIDE=32 HEALPix grid\footnote{https://healpix.jpl.nasa.gov}.
A typical HEALPix cell near the ecliptic has hundreds of asteroid detections per night and even at the
maximum distance from the ecliptic searched, many
asteroids are available.
From this dataset we now have a nightly record of asteroid detections
and non-detections as a function of predicted magnitude within each HEALPix footprint that we can
use to calibrate the detectability of Planet Nine on any night at
any position. Note that we exclusively work in predicted $V$ band magnitude,
ignoring the actual filter used for each observation. This method is equivalent
to assuming that Planet Nine has the same color as the average asteroid.
The effects of other color assumptions are generally small and discussed in
\citet{2022AJ....163..216B}. BB22 examined the use of predicted asteroids
magnitudes as calibrators and found that there were no systematic offsets between
the asteroid predictions and measured brightnesses, and that the RMS deviation
between predictions and measurements is 0.2 magnitudes. We thus conclude that our
self-calibration with asteroids is accurate to approximately this level.
Note that these asteroid extractions are performed after
all of the data filtering discussed above, thus they correctly account for
any real detections that would have been inadvertently removed in the filtering.
\begin{figure*}
\includegraphics[scale=.6]{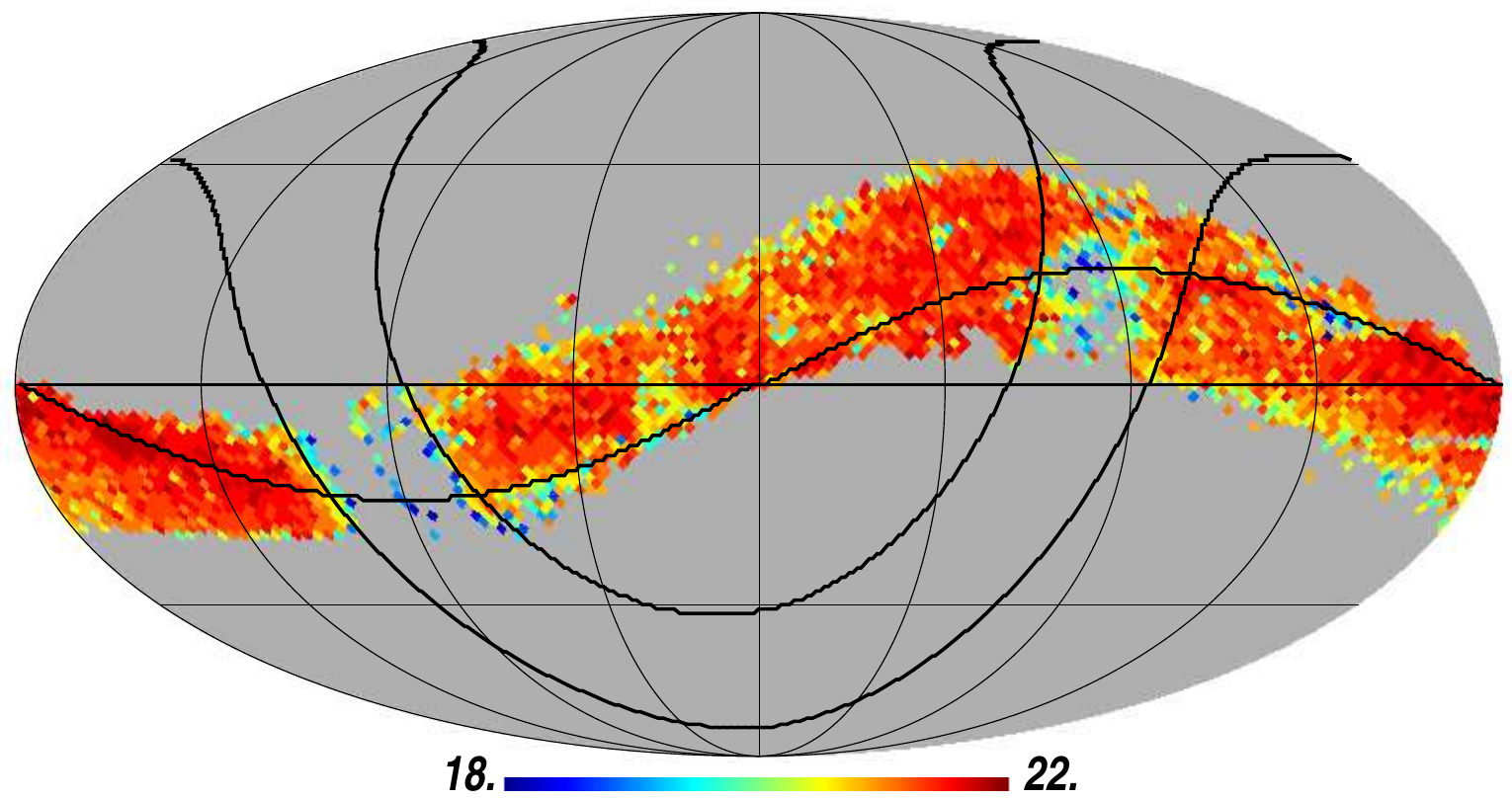}
\caption{The $V$-band magnitude at which there is a 95\% or higher probability
that a moving object would be detected 9 or more times in the
{ portion of the PS1 data that intersects the predicted
locations of P9.}
The data are shown in a Mollweide equal area projection
in equatorial coordinates. Right ascension of 360 is on the left
with 180 in the middle and 0 on the right. The ecliptic
is indicated by a line, as well as galactic latitudes of
$\pm15^\circ$.}
\end{figure*}

We next make use of the Planet Nine reference population of BB22. BB22
created a sample of 100000 potential Planets Nine drawn from the
statistical model of BB21 for the orbital parameters and mass
of Planet Nine. 
They 
assumed a simple mass-diameter relationship of 
$r_9 = (m_9/3) R_{\rm earth}$, where $m_9$ is in Earth masses and $R_{\rm earth}$ is 
the mass of the Earth, based on fits
to planets in this radius and mass range \citep{2013ApJ...772...74W},
and assume 
albedos from half that of Neptune, 0.2, to
a value predicted by a model in which all absorbers
are condensed out of the atmosphere \citep{2016ApJ...824L..25F}, 0.75.
Each Planet Nine was assigned a mean anomaly on
a reference date of 1 June 2018, so their position,
distance from the sun, and brightness can be predicted for 
any night. For each member of this
reference population, we calculate the position 
of the the body for each night of the PS1 survey, determine
the apparent magnitude of the body (ignoring the small
contribution from phase effects), and calculate the
HEALPix grid point in which it would appear. We then use
the record of asteroids detected and not detected on that
night in that grid point to determine the probability that
Planet Nine with its predicted magnitude would have been detected on that night. 
We then randomly select a number between 0 and 1 and, if that number
is lower than the detection probability, we record a detection
of a member of the reference population at that position
-- with an astrometric offset randomly applied based on
the reported uncertainties of asteroids of the same magnitude --
and magnitude.
Note that we do not explicitly consider whether
Planet Nine would be detected in a specific exposure
on a night, but just whether it would have been
detected on any exposure that night and thus result in
a transient object in our PS1 database.
We embed these reference population detections into our PS1 data and use
them for the ultimate calibration of the survey.

The PS1 survey is extremely sensitive to the predicted range of potential 
Planets Nine. Of the 100000 members of the reference population, 88736 would be
detected at least once, with 69082 detected nine times or more. The variable night-to-night detection limit 
at the faint end is ultimately responsible for the
stochastic nature of detections of the faintest
members of the population (as would be the case for
detection of the real Planet Nine at this magnitude).

\section{Orbit linking}
Examining 244 million detections of transient 
objects made over a 5 year period to
find any set of objects
consistent with Keplerian motion is a computational
intensive task. A variation of the 
algorithm developed by \citet{2018AJ....156..135H} and implemented
in BB22 greatly speeds this process. The process, described in detail
in BB22, begins with the simplifying realization that, when viewed from the sun, Keplerian orbits travel in simple great circles across the sky. At the large
distance of Planet Nine, the motions are essentially a constant velocity over long time spans.

\begin{figure*}
\includegraphics[scale=.6]{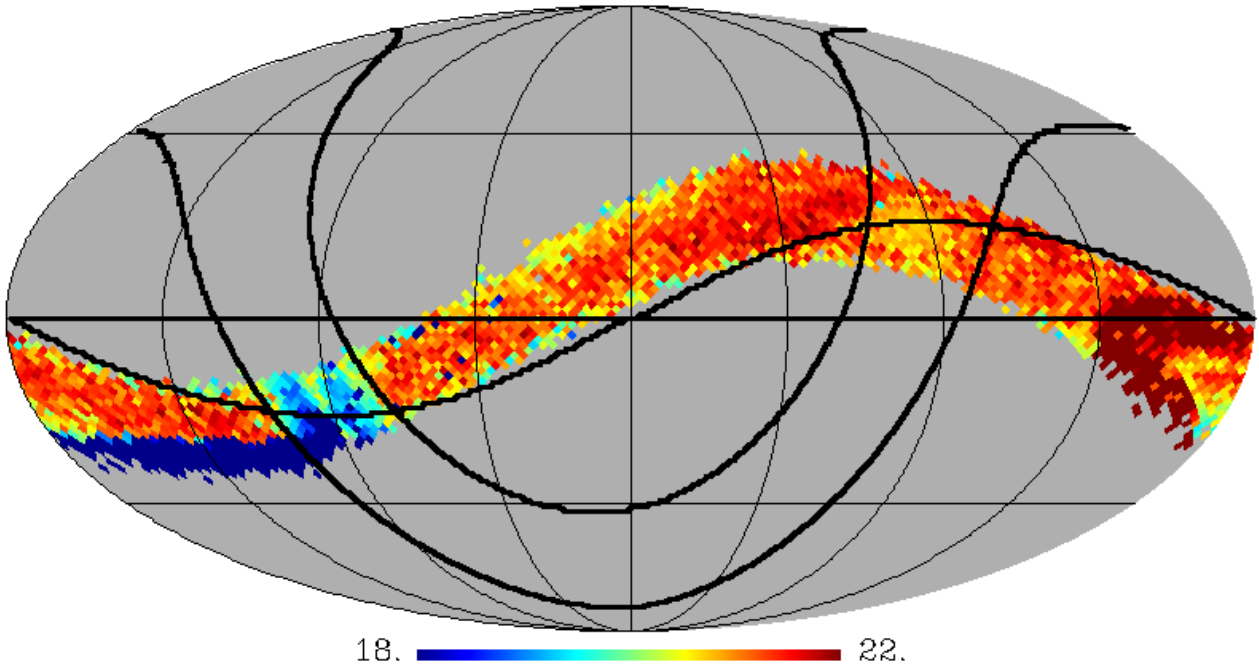}
\caption{The combined $V$-band magnitude limits of the
ZTF, DES, and PS1 suerveys for Planet Nine,
reconstructed from detections of the
synthetic reference population. The
geometry is the same as Fig. 1. Note that the 
sky area is smaller than in Fig. 1 because we
require a minimum of 4 members of the reference
population to be present to estimate a limit.
The deep DES survey on the far right has
magnitude limits that extend as far as 
24. The areas in dark blue remain unsurveyed.}
\end{figure*}
To determine if a transient object is part of a collection
of objects consistent with a Keplerian orbit,
the algorithm takes the object,
assumes a range
of heliocentric distances for this object, and transforms
all other detected transient objects
from their observed
geocentric RA and dec to their heliocentric longitude and
latitude as if they were at the assumed
heliocentric distance and observed from the sun. 
Any real detection of a
Solar System body on a distant Keplerian orbit will now appear as
a collection of transient objects on a great circle at different dates,
separated by a constant
angular speed between the detections. 
To search for such a collection, angular
velocity vectors are calculated from the initial transient object
to every other transient object in the heliocentric system. 
A real detection will now appear as a cluster of objects
with similar angular velocity vectors. A small spread 
in angular velocity vectors occurs even for a real detection
owing to the discrepancy between the assumed and true
distance of the object, but can also occur because the collection
of objects is spurious and due to chance and
does not precisely conform to a
Keplerian orbit.
Thus for every cluster of objects (with a cluster size larger 
than a given number threshold, discussed below),
the cluster of objects (plus the original object)
is fit to a full Keplerian orbit with the 
algorithm of \citet{2000AJ....120.3323B}, and the astrometric
residuals are determined. If the residuals are below
a given threshold, the set of objects
is retained as a candidate for a detection of a real object
with Keplerian motion.

For the PS1 data, we implement this algorithm on the latitudinal
swath
of sky that contains 99\% of the reference population at each
longitude. We further restrict the analysis to motions and distances consistent with this reference population. Distant 
objects which exist but do not fit the Planet Nine hypothesis
will therefore not be found in this analysis. Faster linking
algorithms or increased processing power will be required
before such a larger analysis is possible.

Several choices must be made for the analysis algorithm,
including the spacing of assumed heliocentric distances,
the size of an angular velocity box to be used to identify clusters of
potentially linked objects, the threshold for the minimum 
number of objects to consider a link, and the astrometric
threshold to retain the linkage as a true candidate.
These choices involve a complex set of trade offs.
For example,
a wider spacing of assumed heliocentric distances leads to fewer
geometric transforms but also to the
need to use a larger box in angular velocity 
for identifying clusters, as the
discrepancy between the true and assumed distance causes the
spread of angular velocities to increase.
The larger angular velocity box causes more spurious clusters which
must be checked with the full Keplerian fitting,  drastically
slowing the search.
The most 
efficient trade off is a function of the number density
of detections on the sky.
Similarly,
the threshold for the number of objects required to
be within a cluster before full Keplerian orbit fitting
is a critical parameter.
Requiring a large number of objects to be clustered 
before considering the cluster for Keplerian fitting
greatly speeds
the processing speed at the expense of potentially missing
the real Planet Nine if it is detected a smaller number of times. 
A lower threshold, in contrast, is computationally intensive and also leads to larger numbers
of false positive linkages.

In all cases we choose these parameters in the same manner as
they were chosen in BB22, by simulating different 
spacing for our assumed distances and 
different sizes for our 
angular velocity cluster box sizes in an attempt
to minimize the processing time. While BB22 recalculated
parameters at each location on the sky, here we simplify the 
analysis and use a single angular velocity cluster box width of 0.052 arcseconds day$^{-1}$ in longitude and 0.026 arcseconds day$^{-1}$ in
latitude along with a constant spacing of our assumed distances
in inverse 
heliocentric distance of $\Delta(1/r) = 10^{-5}$ AU$^{-1}$. 
We empirically
find that these parameters come close to optimizing processing
time while, as will be demonstrated below, also finding all
possible linkages.

The final parameter to be selected is the minimum number
of transient detections to be required to be considered
a linked Keplerian orbit. In BB22 we required 7 
detections over a 3 year period. The PS1 data has
a significantly higher number density of (mostly spurious)
objects on the sky,
such that if we require only 7 detections we are overwhelmed
with false linkages. We find that requiring 9 detections
both improves the processing speed and brings the number
of false positives to an acceptably low number.

With these parameters in place we can now visualize approximate
limiting magnitudes for the possible detection of P9 in the PS1
survey. We use our asteroid database to calculate the 
magnitude at which an object would be detected  
on 9 or more distinct dates in a HEALPix grid cell 95\% of the time (Fig. 1). 
Assuming that our processing can efficiently link all such
objects, the median magnitude limit for the detection of
Planet Nine for a region 
outside of the galactic plane is $V=21.0$. 
The northern galactic plane region
has some area of coverage but has considerably worse 
limits, while the southern galactic plane region has almost no
usable data.

\section{Results}
The processing of the data was performed in $3^\circ \times 3^\circ$
blocks across the sky. Enough overlap was included among the blocks
to account for the fastest potential motion of P9 across 
the 5 year period of the data. Even with approximately 50
blocks running in parallel the full data set required several
months of continuous processing time. 

The catalog of simulated reference population detections
was included in the processing,
which had no knowledge whether the detection was from the real
PS1 data or an artificially injected member of the reference
population. Of the 69082 members of the reference population
which had 9 or more detections, 68550 are correctly linked,
for a success rate of 99.2\%. We consider this a strong 
demonstration that, if a Planet Nine candidate consistent with
the 
predicted parameters of BB21
existed in the PS1 data set
and was detected 9 times or more, it would be efficiently
discovered by our algorithm.

In the full dataset, 909 additional linkages are made. 
Each one of them is a chance linkage between multiple
members of the reference population and a small
number of fortuitously placed real PS1 detections (note that
the algorithm allows multiple linkages between objects, so in 
each case the true linkage of the objects belonging to the
reference population is still correctly made). 

We conclude that no object with the predicted characteristics
of P9 was detected nine or more times 
in the PS1 data set. The magnitude limits shown in Fig. 1
provide a good approximation to the search limit, with the
caveat that the search is only sensitive to objects with orbital
characteristics similar to those predicted by BB21.

The reference population provides an effective method for
combining the limits from the PS1 data with those already
achieved by ZTF and DES. Of the 69802 members of the
reference population detected, 17054 were unique to PS1.
The total fraction of the reference population that has been
ruled out by the combination of ZTF, DES, and PS1 is 
78\%. 
In Figure 2 we show an approximation of the
full magnitude limit of the three combined
surveys by examining each HEALPix grid point
that contains 4 or more members of the reference
population and setting the limit equal to
the faintest object detected brighter than
the brightest non-detection. When all objects
in the grid point are detected we show the
magnitude of the faintest object. 
Each of the surveys has regions of unique
contribution. PS1 uniquely detects
17\% of the reference population, mostly
at high galactic latitudes at depths fainter
than ZTF. ZTF uniquely detects 6\%,
predominantly in the northern galactic plane
where PS1 coverage is poor. And DES uniquely 
detects 3\% of the population at magnitudes
fainter than PS1 covers. The remaining 52\%
of the detected population is detected by
two or more surveys, most often ZTF+PS1 for
the brighter objects at high galactic latitudes.

\begin{figure}
    \centering
    \includegraphics[scale=.5]{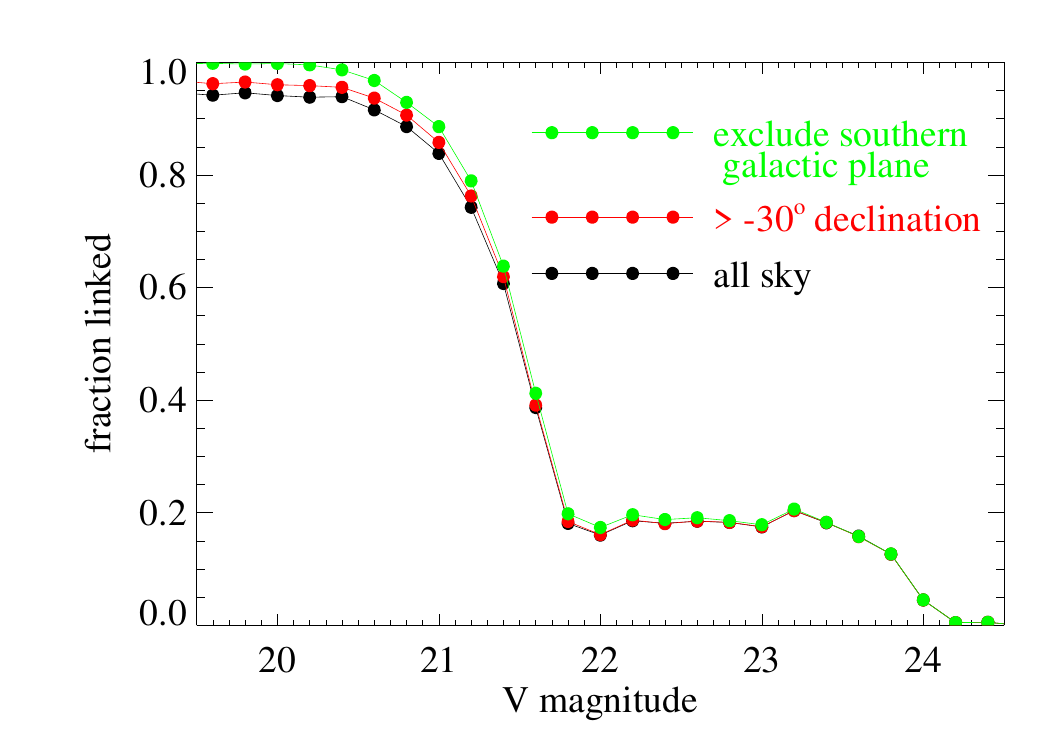}
    \caption{The fraction of the BB22 Planet Nine reference population that would have been
    detected by the ZTF, DES, and PS1 surveys, as a function of $V$-band magnitude. The combined
    surveys are extremely efficient for objects fainter than $V\sim 21$ and reach 50\% efficiency at $V=21.5$, with a large tail of fainter detections from
    the deep but narrow DES survey.}
\end{figure}

As is apparent, the combined surveys have a 
step-wise efficiency with magnitude (Fig. 3).
The combined survey is about 94\% efficient
for objects brighter than $V=20.5$, falling
to a 50\% efficient at $V=21.5$. A large
tail of detectable objects as faint as $V=23.5$, exclusively
from the DES observations, extends at $\sim$20\%
efficiency.
Many of
the missing bright objects are in the
unobserved low declination regions or the
poorly observed southern galactic plane. If these regions
are excluded, the bright object
efficiency increases to 97\% and 99.7\%,
respectively. There is reason to believe
that P9 would not be found at these locations:
a full fit of 
planetary ephemerides and search for
gravitational perturbations due to P9
suggests that a P9
near perihelion -- as these southern
hemisphere positions are -- would have 
already been detected \citep{2020A&A...640A...6F}. 
How best to combine these constraints with those
derived here remains uncertain, however.

\begin{figure*}
    \centering
    \includegraphics[scale=.6]{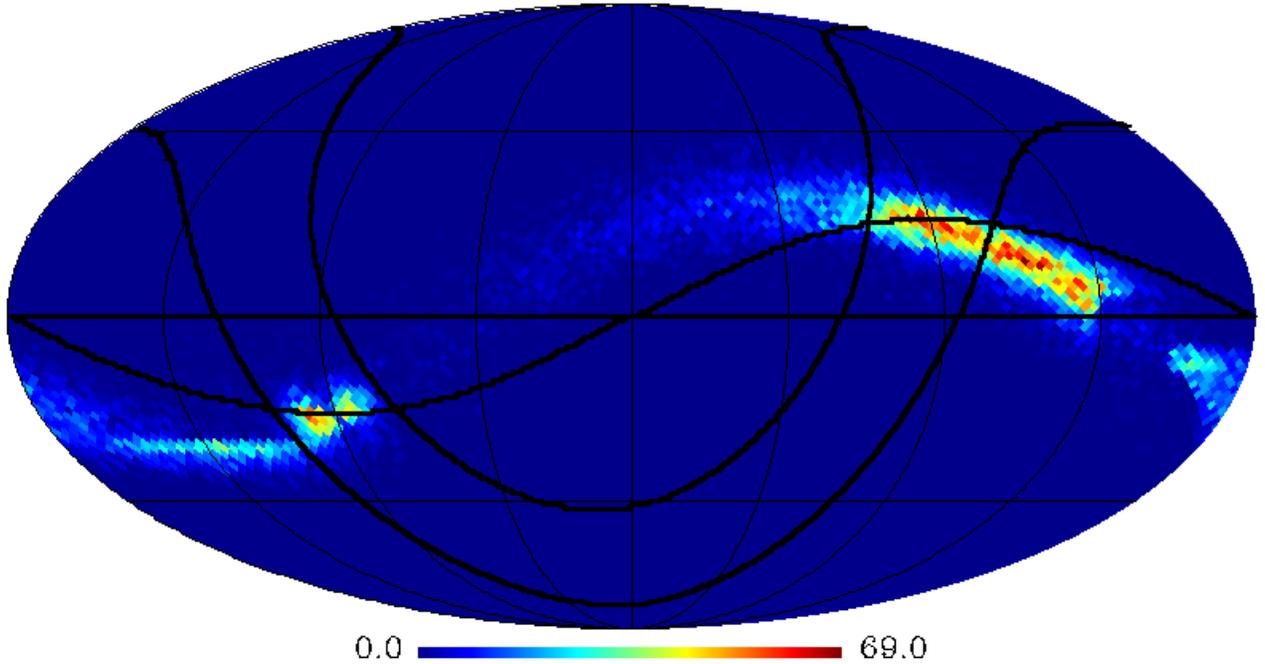}
    \caption{The probability density function of on-sky location of the BB22 Planet Nine reference
    population that would remain undetected after the ZTF, DES, and PS1 surveys. The geometry is the
    same as Figure 1.}
\end{figure*}
\section{Discussion}
The combination of the ZTF, DES, and PS1
surveys rules out 78\% of the BB22 P9 reference population.
The sky positions of the remaining members of the Reference
{ Population are shown in Figure 4.}
Using the members
of the Reference Population yet to be ruled out,
we update our predictions
for P9 parameters. 
We report the median with the 15.8 
and 84.1 percentile values as the uncertainties.
Our newly updated estimates include
a semimajor axis of $500^{+170}_{-120}$ AU,
a mass of $6.6^{+2.6}_{-1.7}$ M$_{\earth}$,
an aphelion distance of $630^{+290}_{-170}$ AU,
a current distance of 550$^{+250}_{-180}$ AU,
and a V magnitude of 22.0$^{+1.1}_{-1.4}$.
The other predicted parameters remain generally
unchanged. All parameters and distributions
of the Planet Nine reference population,
including flags for whether the objects
would have been found by ZTF, by DES, or by PS1,
can be found permanently archived at 
\url{https://data.caltech.edu/records/8fjad-x7y61}

The remaining areas of P9 parameter space that remain unexplored,
shown in Figure 4, include a large swath near the 
northern galactic plane where P9 is near aphelion and thus would be fainter than the current limits
and a smaller region in the part of the southern galactic plane that remains poorly covered as
well as a strip below a declination of -30$^{\circ}$.
{ Much of this remaining parameter space will be covered by the upcoming Vera Rubin Observatory
survey, which will be sensitive to all but the faintest and most northern predicted positions.}

While a large majority of the phase space for Planet Nine predicted by
BB21 has now been ruled out, significant 
regions still remain unobserved to the needed depth. Nonetheless, it is worth considering
potential reasons why P9 was not found in the first 78\% of parameter space surveyed.
{ An obvious possibility, of course, is that Planet Nine does not exist. 
Such an explanation would require new explanations for multiple phenomena observed
in the outer Solar System. Until such explanations are available, we continue to regard
Planet Nine as the most likely hypothesis.}
\citet{2022AJ....163..216B} explore the effects of different assumed colors, albedos,
and radii for Planet Nine, and show that different choices of these parameters
can change the amount of phase space covered by only of order $\sim$10\%.
A potentially much larger effect would be a change in source region of the 
objects which become clustered by Planet Nine. In BB21, the assumption is made that the
objects being observed are sourced from an early extended
scattered disk. \citet{2021ApJ...910L..20B}
instead consider the effects of Planet Nine on objects pulled in from the 
inner Oort cloud and conclude that a similar clustering is observed for these objects
but that the width of the cluster is broader. In BB21, the breadth of the cluster 
is directly related to the mass of Planet Nine and the broad
cluster observed in the distant solar system
is used to infer a lower mass, lower semimajor axis Planet Nine.
If, instead, the observed breadth is caused
by an inner Oort cloud source population for the clustered objects, the 
true Planet Nine could be more massive and more distant,
making it potentially much fainter and
harder to find. More work is required to explore this alternative version of the
Planet Nine hypothesis. 
\acknowledgements
The Pan-STARRS1 Surveys (PS1) and the PS1 public science archive have been made possible through contributions by the Institute for Astronomy, the University of Hawaii, the Pan-STARRS Project Office, the Max-Planck Society and its participating institutes, the Max Planck Institute for Astronomy, Heidelberg and the Max Planck Institute for Extraterrestrial Physics, Garching, The Johns Hopkins University, Durham University, the University of Edinburgh, the Queen's University Belfast, the Harvard-Smithsonian Center for Astrophysics, the Las Cumbres Observatory Global Telescope Network Incorporated, the National Central University of Taiwan, the Space Telescope Science Institute, the National Aeronautics and Space Administration under Grant No. NNX08AR22G issued through the Planetary Science Division of the NASA Science Mission Directorate, the National Science Foundation Grant No. AST-1238877, the University of Maryland, Eotvos Lorand University (ELTE), the Los Alamos National Laboratory, and the Gordon and Betty Moore Foundation.

\bibliography{kbo0421}{}

\begin{thebibliography}{}
\expandafter\ifx\csname natexlab\endcsname\relax\def\natexlab#1{#1}\fi
\providecommand{\url}[1]{\href{#1}{#1}}
\providecommand{\dodoi}[1]{doi:~\href{http://doi.org/#1}{\nolinkurl{#1}}}
\providecommand{\doeprint}[1]{\href{http://ascl.net/#1}{\nolinkurl{http://ascl.net/#1}}}
\providecommand{\doarXiv}[1]{\href{https://arxiv.org/abs/#1}{\nolinkurl{https://arxiv.org/abs/#1}}}

\bibitem[{{Babinet}(1848)}]{Babinet1848}
{Babinet}, M. 1848, Comptes Rendus, 27, 202

\bibitem[{{Batygin} \& {Brown}(2016{\natexlab{a}})}]{2016AJ....151...22B}
{Batygin}, K., \& {Brown}, M.~E. 2016{\natexlab{a}}, \aj, 151, 22,
  \dodoi{10.3847/0004-6256/151/2/22}

\bibitem[{{Batygin} \& {Brown}(2016{\natexlab{b}})}]{2016ApJ...833L...3B}
---. 2016{\natexlab{b}}, \apjl, 833, L3, \dodoi{10.3847/2041-8205/833/1/L3}

\bibitem[{{Batygin} \& {Brown}(2021)}]{2021ApJ...910L..20B}
---. 2021, \apjl, 910, L20, \dodoi{10.3847/2041-8213/abee1f}

\bibitem[{{Belyakov} {et~al.}(2022){Belyakov}, {Bernardinelli}, \&
  {Brown}}]{2022AJ....163..216B}
{Belyakov}, M., {Bernardinelli}, P.~H., \& {Brown}, M.~E. 2022, \aj, 163, 216,
  \dodoi{10.3847/1538-3881/ac5c56}

\bibitem[{{Bernardinelli} {et~al.}(2022){Bernardinelli}, {Bernstein}, {Sako},
  {Yanny}, {Aguena}, {Allam}, {Andrade-Oliveira}, {Bertin}, {Brooks},
  {Buckley-Geer}, {Burke}, {Carnero Rosell}, {Carrasco Kind}, {Carretero},
  {Conselice}, {Costanzi}, {da Costa}, {De Vicente}, {Desai}, {Diehl},
  {Dietrich}, {Doel}, {Eckert}, {Everett}, {Ferrero}, {Flaugher}, {Fosalba},
  {Frieman}, {Garc{\'\i}a-Bellido}, {Gerdes}, {Gruen}, {Gruendl}, {Gschwend},
  {Hinton}, {Hollowood}, {Honscheid}, {James}, {Kent}, {Kuehn}, {Kuropatkin},
  {Lahav}, {Maia}, {March}, {Menanteau}, {Miquel}, {Morgan}, {Myles}, {Ogando},
  {Palmese}, {Paz-Chinch{\'o}n}, {Pieres}, {Plazas Malag{\'o}n}, {Romer},
  {Roodman}, {Sanchez}, {Scarpine}, {Schubnell}, {Serrano}, {Sevilla-Noarbe},
  {Smith}, {Soares-Santos}, {Suchyta}, {Swanson}, {Tarle}, {To}, {Varga}, \&
  {Walker}}]{2022ApJS..258...41B}
{Bernardinelli}, P.~H., {Bernstein}, G.~M., {Sako}, M., {et~al.} 2022, \apjs,
  258, 41, \dodoi{10.3847/1538-4365/ac3914}

\bibitem[{{Bernstein} \& {Khushalani}(2000)}]{2000AJ....120.3323B}
{Bernstein}, G., \& {Khushalani}, B. 2000, \aj, 120, 3323,
  \dodoi{10.1086/316868}

\bibitem[{{Brown} \& {Batygin}(2019)}]{2019AJ....157...62B}
{Brown}, M.~E., \& {Batygin}, K. 2019, \aj, 157, 62,
  \dodoi{10.3847/1538-3881/aaf051}

\bibitem[{{Brown} \& {Batygin}(2021)}]{2021AJ....162..219B}
---. 2021, \aj, 162, 219, \dodoi{10.3847/1538-3881/ac2056}

\bibitem[{{Brown} \& {Batygin}(2022)}]{2022AJ....163..102B}
---. 2022, \aj, 163, 102, \dodoi{10.3847/1538-3881/ac32dd}

\bibitem[{{Brown} {et~al.}(2004){Brown}, {Trujillo}, \&
  {Rabinowitz}}]{2004ApJ...617..645B}
{Brown}, M.~E., {Trujillo}, C., \& {Rabinowitz}, D. 2004, \apj, 617, 645,
  \dodoi{10.1086/422095}

\bibitem[{{Chambers} {et~al.}(2016){Chambers}, {Magnier}, {Metcalfe},
  {Flewelling}, {Huber}, {Waters}, {Denneau}, {Draper}, {Farrow}, {Finkbeiner},
  {Holmberg}, {Koppenhoefer}, {Price}, {Rest}, {Saglia}, {Schlafly}, {Smartt},
  {Sweeney}, {Wainscoat}, {Burgett}, {Chastel}, {Grav}, {Heasley}, {Hodapp},
  {Jedicke}, {Kaiser}, {Kudritzki}, {Luppino}, {Lupton}, {Monet}, {Morgan},
  {Onaka}, {Shiao}, {Stubbs}, {Tonry}, {White}, {Ba{\~n}ados}, {Bell},
  {Bender}, {Bernard}, {Boegner}, {Boffi}, {Botticella}, {Calamida},
  {Casertano}, {Chen}, {Chen}, {Cole}, {Deacon}, {Frenk}, {Fitzsimmons},
  {Gezari}, {Gibbs}, {Goessl}, {Goggia}, {Gourgue}, {Goldman}, {Grant},
  {Grebel}, {Hambly}, {Hasinger}, {Heavens}, {Heckman}, {Henderson}, {Henning},
  {Holman}, {Hopp}, {Ip}, {Isani}, {Jackson}, {Keyes}, {Koekemoer}, {Kotak},
  {Le}, {Liska}, {Long}, {Lucey}, {Liu}, {Martin}, {Masci}, {McLean}, {Mindel},
  {Misra}, {Morganson}, {Murphy}, {Obaika}, {Narayan}, {Nieto-Santisteban},
  {Norberg}, {Peacock}, {Pier}, {Postman}, {Primak}, {Rae}, {Rai}, {Riess},
  {Riffeser}, {Rix}, {R{\"o}ser}, {Russel}, {Rutz}, {Schilbach}, {Schultz},
  {Scolnic}, {Strolger}, {Szalay}, {Seitz}, {Small}, {Smith}, {Soderblom},
  {Taylor}, {Thomson}, {Taylor}, {Thakar}, {Thiel}, {Thilker}, {Unger},
  {Urata}, {Valenti}, {Wagner}, {Walder}, {Walter}, {Watters}, {Werner},
  {Wood-Vasey}, \& {Wyse}}]{2016arXiv161205560C}
{Chambers}, K.~C., {Magnier}, E.~A., {Metcalfe}, N., {et~al.} 2016, arXiv
  e-prints, arXiv:1612.05560.
\newblock \doarXiv{1612.05560}

\bibitem[{{Fienga} {et~al.}(2020){Fienga}, {Di Ruscio}, {Bernus}, {Deram},
  {Durante}, {Laskar}, \& {Iess}}]{2020A&A...640A...6F}
{Fienga}, A., {Di Ruscio}, A., {Bernus}, L., {et~al.} 2020, \aap, 640, A6,
  \dodoi{10.1051/0004-6361/202037919}

\bibitem[{{Fortney} {et~al.}(2016){Fortney}, {Marley}, {Laughlin},
  {Nettelmann}, {Morley}, {Lupu}, {Visscher}, {Jeremic}, {Khadder}, \&
  {Hargrave}}]{2016ApJ...824L..25F}
{Fortney}, J.~J., {Marley}, M.~S., {Laughlin}, G., {et~al.} 2016, \apjl, 824,
  L25, \dodoi{10.3847/2041-8205/824/2/L25}

\bibitem[{{Holman} {et~al.}(2018){Holman}, {Payne}, {Blankley}, {Janssen}, \&
  {Kuindersma}}]{2018AJ....156..135H}
{Holman}, M.~J., {Payne}, M.~J., {Blankley}, P., {Janssen}, R., \&
  {Kuindersma}, S. 2018, \aj, 156, 135, \dodoi{10.3847/1538-3881/aad69a}

\bibitem[{{Madigan} \& {McCourt}(2016)}]{2016MNRAS.457L..89M}
{Madigan}, A.-M., \& {McCourt}, M. 2016, \mnras, 457, L89,
  \dodoi{10.1093/mnrasl/slv203}

\bibitem[{{Napier} {et~al.}(2021){Napier}, {Gerdes}, {Lin}, {Hamilton},
  {Bernstein}, {Bernardinelli}, {Abbott}, {Aguena}, {Annis}, {Avila}, {Bacon},
  {Bertin}, {Brooks}, {Burke}, {Carnero Rosell}, {Carrasco Kind}, {Carretero},
  {Costanzi}, {da Costa}, {De Vicente}, {Diehl}, {Doel}, {Everett}, {Ferrero},
  {Fosalba}, {Garc{\'\i}a Bellido}, {Gruen}, {Gruendl}, {Gutierrez},
  {Hollowood}, {Honscheid}, {Hoyle}, {James}, {Kent}, {Kuehn}, {Kuropatkin},
  {Maia}, {Menanteau}, {Miquel}, {Morgan}, {Palmese}, {Paz-Chinch{\'o}n},
  {Plazas}, {Sanchez}, {Scarpine}, {Serrano}, {Sevilla-Noarbe}, {Smith},
  {Suchyta}, {Swanson}, {To}, {Walker}, \& {Wilkinson}}]{2021arXiv210205601N}
{Napier}, K.~J., {Gerdes}, D.~W., {Lin}, H.~W., {et~al.} 2021, arXiv e-prints,
  arXiv:2102.05601.
\newblock \doarXiv{2102.05601}

\bibitem[{{Nesvorn{\'y}} {et~al.}(2023){Nesvorn{\'y}}, {Bernardinelli},
  {Vokrouhlick{\'y}}, \& {Batygin}}]{2023Icar..40615738N}
{Nesvorn{\'y}}, D., {Bernardinelli}, P., {Vokrouhlick{\'y}}, D., \& {Batygin},
  K. 2023, \icarus, 406, 115738, \dodoi{10.1016/j.icarus.2023.115738}

\bibitem[{{Scholtz} \& {Unwin}(2020)}]{2020PhRvL.125e1103S}
{Scholtz}, J., \& {Unwin}, J. 2020, \prl, 125, 051103,
  \dodoi{10.1103/PhysRevLett.125.051103}

\bibitem[{{Sefilian} \& {Touma}(2019)}]{2019AJ....157...59S}
{Sefilian}, A.~A., \& {Touma}, J.~R. 2019, \aj, 157, 59,
  \dodoi{10.3847/1538-3881/aaf0fc}

\bibitem[{{Shankman} {et~al.}(2017){Shankman}, {Kavelaars}, {Bannister},
  {Gladman}, {Lawler}, {Chen}, {Jakubik}, {Kaib}, {Alexandersen}, {Gwyn},
  {Petit}, \& {Volk}}]{2017AJ....154...50S}
{Shankman}, C., {Kavelaars}, J.~J., {Bannister}, M.~T., {et~al.} 2017, \aj,
  154, 50, \dodoi{10.3847/1538-3881/aa7aed}

\bibitem[{{Sivaram} {et~al.}(2016){Sivaram}, {Kenath}, \&
  {Kiren}}]{2016Ap&SS.361..230S}
{Sivaram}, C., {Kenath}, A., \& {Kiren}, O.~V. 2016, \apss, 361, 230,
  \dodoi{10.1007/s10509-016-2815-z}

\bibitem[{{Standish}(1993)}]{1993AJ....105.2000S}
{Standish}, E.~M. 1993, \aj, 105, 2000, \dodoi{10.1086/116575}

\bibitem[{{Tonry} {et~al.}(2018){Tonry}, {Denneau}, {Flewelling}, {Heinze},
  {Onken}, {Smartt}, {Stalder}, {Weiland}, \& {Wolf}}]{2018ApJ...867..105T}
{Tonry}, J.~L., {Denneau}, L., {Flewelling}, H., {et~al.} 2018, \apj, 867, 105,
  \dodoi{10.3847/1538-4357/aae386}

\bibitem[{{Trujillo} \& {Sheppard}(2014)}]{2014Natur.507..471T}
{Trujillo}, C.~A., \& {Sheppard}, S.~S. 2014, \nat, 507, 471,
  \dodoi{10.1038/nature13156}

\bibitem[{{Wu} \& {Lithwick}(2013)}]{2013ApJ...772...74W}
{Wu}, Y., \& {Lithwick}, Y. 2013, \apj, 772, 74,
  \dodoi{10.1088/0004-637X/772/1/74}

\end{thebibliography}
\bibliographystyle{aasjournal}

\end{document}